# Demonstration of Thermal Images Captured by a Backside Illuminated Transition Edge Bolometer

Roya Moahjeri[1], Seyed Iman Mirzaei[2], Rana Nazifi[1], Anders Christian Wulff[3], Jean-Claude Grivel[3], and Mehdi Fardmanesh[1]

[1] Department of Electrical Engineering, Sharif University of Technology, Tehran, Iran
[2] Faculty of Basic Sciences, Tarbiat Modares Unievrsity, Tehran, Iran
[3] Department of Energy Conversion and Storage, Technical University of Denmark, 2800 Kgs Lyngby, Denmark

Corresponding email: fardmanesh@sharif.edu

**Abstract**

A high temperature superconducting detector was fabricated to capture the thermal images in room temperature background. The detector was made of YBa2Cu3O7-δ (YBCO) superconducting thin film deposited on an Yttria Stabilized Zirconia (YSZ) substrate, structurally modified for high responsivity performance. A very thin absorber layer was implemented on the backside of the detector to increase the absorption of the device at infrared (IR) range of frequencies. To capture thermal images, a movable off-axis parabolic mirror on XY plane was used for focusing the incoming IR radiation including the thermal signal of the objects onto the device surface. The captured thermal images belong to the objects within the temperature range between 300K and 400K, which correspond to our imaging optics.

## 1. Introduction

IR detectors capture the emitted infrared radiation from objects with temperatures above absolute zero, which have found great utility in industrial, medical, and military applications [1,2]. These detectors are generally classified to the quantum type that operate at very low temperatures in a narrow range of frequencies with ultra-high sensitivity [3,4] and the thermal type with the possibility of operating at higher temperatures, well suited for broadband detection [4]. Bolometers are a particular type of thermal detectors that operate based on the changes in their electrical resistance upon the absorption of radiation power [5]. Since there is a great flexibility in choosing the absorbing material, bolometers can readily be designed to have narrow or broad-band characteristics over a wide range of wavelengths.

Superconducting transition edge bolometers (TEB) are known as highly sensitive bolometers due to the sharp temperature variation of their resistance (dR/dT) within the normal to superconductor transition width, which is controllable by implementation of different processing parameters [6,7]. To get higher sensitivity from the TEB detectors, they are thermally biased at the middle of the transition region, where the maximum dR/dT is achievable. Therefore, if the device is electrically biased by a constant current, the change in the voltage of the bolometer is a demonstration of the incoming radiation power. The voltage response behavior of these kind of bolometer detectors to a chopped radiation at low frequencies, was formulated by a one-dimensional thermophysical model defined as [8]:

$$r_v = \frac{dV_r}{dP} = \frac{\eta I}{G + j2\pi f C}\frac{dR}{dT}, \qquad (1)$$

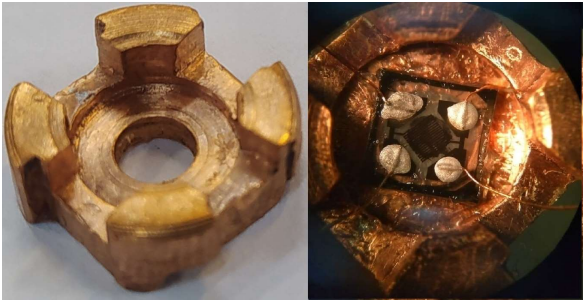

Figure 1. Left) the copper holder, right) the fabricated bolometer mounted on the copper holder.

where $V_r$ is the voltage response of the bolometer, P is the incoming radiation power, η is the absorption coefficient of the detector, I is the DC bias current, R is the resistance of the bolometer, G is the thermal conductivity between the device and the temperature reservoir, C is the heat capacity of the substrate, and f is the modulation frequency of the incoming radiation. According to (1), higher dR/dT and lower thermal conductivity and heat capacity lead to an enhanced voltage responsivity.

In this work, we demonstrate thermal images, captured by a fabricated broad-band single pixel transition edge bolometer (TEB). The TEB detector is made of YBCO superconducting thin film on an Yttria Stabilized Zirconia (YSZ) substrate using a low-cost fabrication technology, which facilitates its application in arrayed structures with large number of integrated pixels [9]. The critical temperature of the superconducting film is about 90K with the transition width of less than 2K. To increase the absorption of the TEB detector at IR frequencies, a very thin absorber layer was implemented at the backside of the device, on which the incoming radiation was directed by an off axis parabolic mirror. The mirror is mounted on a XY scanner, which was controlled using stepper motors, drivers and Arduino microcontroller that is connected to the PC and controlled by Python code.

## 2. Fabrication of the thermal detector

An Yttrium Stabilized Zirconia (YSZ) crystal was used as a substrate for the deposition of YBCO superconducting thin film using Metal organic deposition technique. To acquire an appropriate lattice match between the substrate and YBCO thin film, a $Ce_{0.9}La_{0.1}O_2$ (CLO) buffer layer with a thickness of about 20 nm is deposited on the YSZ substrate using a precursor solution-based method [10]. Subsequently, the YBCO film was patterned to meander line structure, on either sides of which two contact pads were introduced, necessary for four point probe responsivity measurement. The fabricated device mounted on a copper holder is shown in figure 1.
According to equation (1), lowering the thermal conductivity and the heat capacity leads to higher voltage responsivity. To decrease the thermal capacitance of the detector, the substrate is polished from the backside reaching a thickness of about 200 μm. Then an IR absorber was implemented on the backside of the device to collect radiation energy power in the form of heat on its surface, while avoiding the negative effect of the absorber layer on the quality of the superconducting thin film [11]. The absorber layer thickness is very thin to decrease its thermal capacitance compared to that of the substrate. The configuration of the device in thermal contact with a copper holder is shown in figure 2.

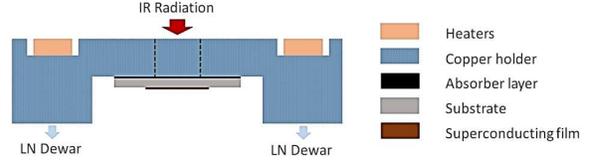

Figure 2. The configuration of the device mounted on the copper holder with a hole at its center. The holder is connected to LN dewar from the bottom.

According to figure 2, the copper holder is connected to a liquid nitrogen (LN) based dewar with a vacuum encapsulation to reduce thermal fluctuations due to the convective heat transfer. To thermally bias the TEB at the middle of the superconducting transition region, two heaters were uniformly placed on the cold head to provide a homogeneous temperature distribution over its volume [8, 12]. Two sensors were perpendicularly placed on the holder for temperature readout, which is not visible in figure 2. As shown in figure 2, the device is exposed to the incoming IR radiation from the absorber side, freely standing on the copper holder. This freestanding structure improves the detector voltage responsivity at low modulation frequencies due to a reduced thermal conductivity.

## 3. The thermal imaging system

The thermal imaging setup is mainly composed of the radiation detector, detector readout electronics, home-made optics and computer controlled XY stage. TEB sensor detects the heat energy being emitted by an object and converts it into an electrical signal, which is measured by the readout electronics. The readout electronics and the imaging optics are controlled and synchronized by a python script on a PC. The imaging optics is mainly composed of a home-made off-axis parabolic mirror with a focal length of around 10 cm, designed to get less distorted image during scanning. The mirror is mounted on a XY scanner, which is controlled using Arduino board controlled by Python code. The image is reconstructed by correlating the acquired data to the position information. The configuration of the imaging setup is shown in fiqure 3a. Figure 3b shows the LN dewar required for the operation of our TEB detector with a mechanical chopper standing in front of it. The parabolic mirror on the movable XY stage is also displayed in figure 3c.

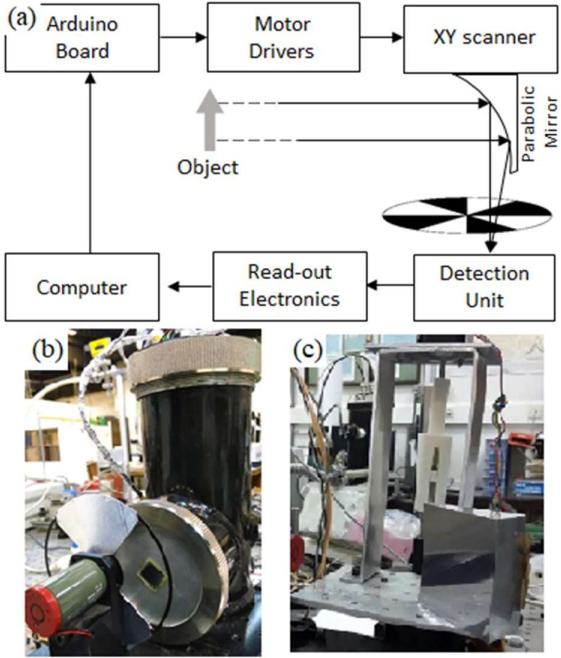

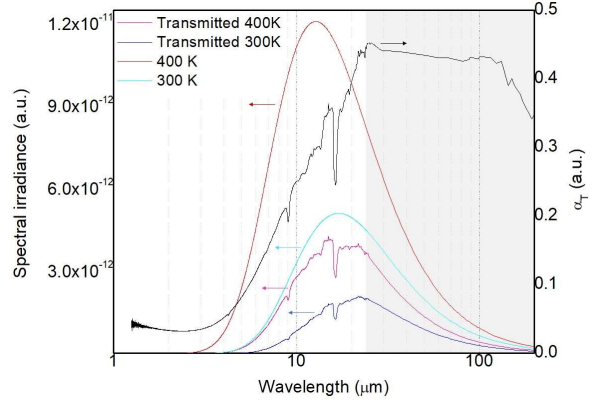

Figure 4. The transmission coefficient of the silicon window (black), Blackbody radiation of an object with temperatures of 400K, 300K respectively before (red), (light blue), and after (pink), (dark blue) passing through the silicon window.

Figure 3. a) Demonstration of the thermal imaging setup, b) LN dewar with the mechanical chopper in fornt, c) parabolic mirror on XY stage.

## 4. Results and Discussion

The information about the temperature of an object mainly resides in the IR range of frequencies, which has to get selectively detected. Therefore, first of all the IR radiation should enter the dewar to impinge on the detector surface, and then the detector should be able to collect it. Here, a 2cm×2cm uncoated and undoped silicon wafer with a thickness of 500 µm was used as the IR-transparent window of the LN dewar. To measure its transmission spectrum, a Bruker Vertex 70 spectrometer has been utilized in the wavelength range between 1.5µm and 25µm, and from 25µm to 200µm, a globar based FTIR custom-made spectrometer was used.

According to figure 4, at the wavelengths between 2.5µm up to 25 µm, an increase was observed in the the silicon transmission from 0.1 up to 0.45. Above the wavelength of 25µm, a slight change in this transmission is percieved up to 200µm. Due to this variation, a distortion is expected in the transmitted spectrum of a blackbody radiation through the silicon window, and hence the amount of the detected power by the TEB detector. Moreover, the peak of the radiation curve moves to lower wavelengths as the temperature of the blackbody increases, which is less transmitted by the window. The blackbody curve at 400K and 300K, and the pertinent transmitted spectrum are shown in figure 4. Using Stefan–Boltzmann law [13], it can be calculated that the absolute temperatures of 400K and 300K is attenuated to 283K and 223K respectively by passing through the window. Therefore, the dynamic range of the imaging system is influenced by the properties of the silicon window in wider temperature ranges.

To increase the sensitivity of the detector to IR range of frequencies, it is necessary to use an absorber layer on the detector. Covering the YBCO film with an absorber layer destructively influences the electrical properties of the superconducting YBCO material [11]. Therefore, the absorber layer was implemented on the backside of the substrate. Here, black marker and carbon black were utilized as the absorber layers, of which the measured absorption coefficients are shown in figure 5. According to figure 5, the absorption coefficient of the absorber layers are considerably higher than that of the YBCO film, which is highly reflective in IR range of frequencies [14]. This is while he effective absorbing area has more than doubled by using the absorber layers, which further improves the sensitivity of the bolometer. The thickness of the implemented absorber layers was kept very low to get a negligible heat capacity compared to that of the substrate.

As it is demonstrated in figure 2, the detector is excited from the backside, opposite to the YBCO grown side of the substrate. Upon the absorption of the radiation, the generated heat at the surface of the absorber is propagated towards the YBCO film through the substrate. Then the resistance of the bolometer changes, and the voltage response across this resistance is measured by a low noise electronics, which is shown in figure 6. To limit the measurement noise to that of the chopping frequency, the incoming radiation is modulated by a mechanical chopper. Therefore, the heat diffusion length into the substrate is determined as [8]:

$$L_f = \sqrt{\frac{D}{\pi f}}, \qquad (2)$$

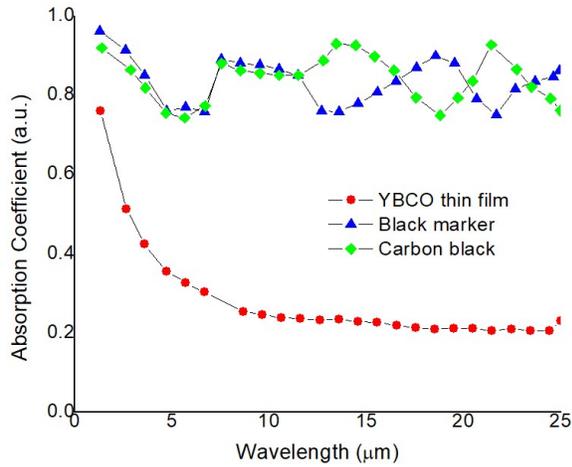

Figure 5. The absorption coefficient of YBCO thin film, Black marker and carbon black.

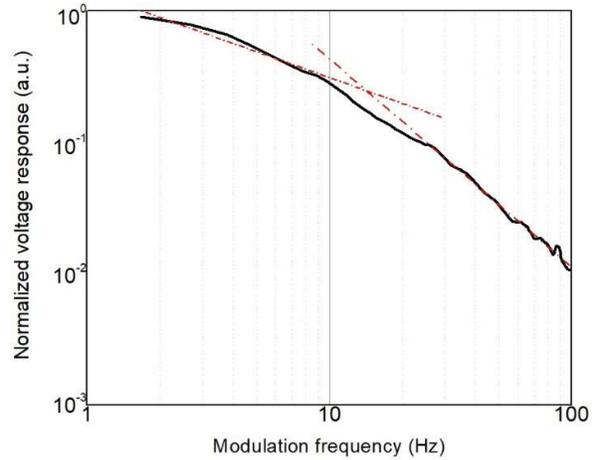

Figure 6. The Normalized voltage response of the fabricated bolometer versus modulation frequency with the intersection point between two tangent lines to the curve showing the modulation frequency limit.

where, D=K/C is the thermal diffusivity of the substrate. K, and C are the thermal conductivity and the specific heat capacity of the substrate material, respectively.

According to (2), the thermal diffusion length is inversely proportional to the modulation frequency, which means that above a certain frequency the voltage responsivity of the bolometer will decrease with larger slope. Having the thermal parameters of the YSZ substrate [15], this frequency limit for our detector with the substrate thickness of 200μm is calculated to be around 13 Hz. This frequency limit is also observable in figure 6, specified by the intersection point between two lines defining the linear approximation for the curve. Therefore, there is a tradeoff between the voltage response amplitude, and hence the sensitivity of the detector and the detection speed. In fact, at higher modulation frequencies, the amplitude of the voltage response is relatively lower with a faster detection speed. A modulation frequency of 10 Hz was used for the imaging measurements, corresponding to the voltage responsivity characteristics of the bolometer.

Thermal images captured by the fabricated bolometer using the imaging setup in section 3, are shown in figure 7, visualized linearly in "hot" colormap. The distances from the objects to the mirror were about 5 meters, where the objects could be considered essentially infinitely far away, compared to 10cm focal length for the parabolic mirror. Therefore, the parallel incident IR rays including the thermal signal of the objects were focused at the focal point of the mirror, where the detector was located. As it is observed in figure 7, there is a shadow around the captured images, which is attributed to the gradual increase of the background radiation contribution in the parallel incident IR rays by approaching to the edges of the object image through scanning.

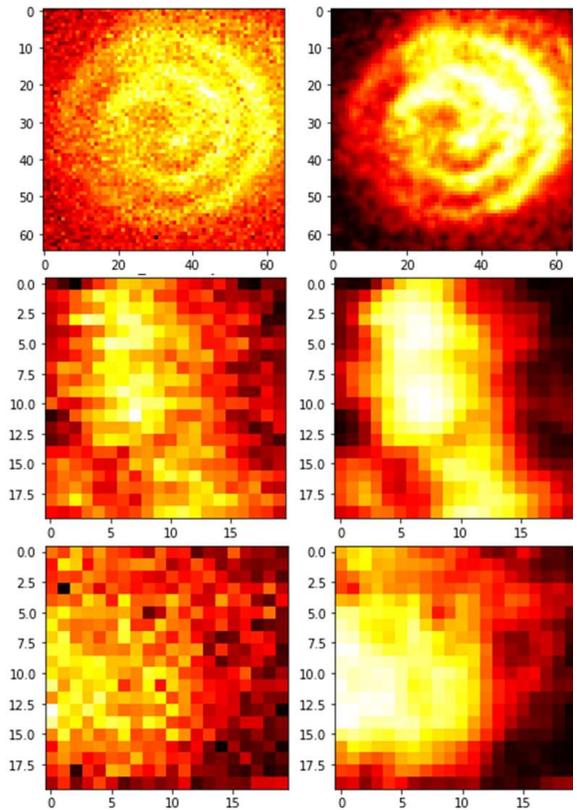

Figure 7. From top to bottom respectively, thermal images of a current driven spiral wire, the human face, and a cup full of 100ºC water are displayed. The points of the objects with the highest and the lowest temperatures are mapped to the white and black in the "hot" colormap.

The first image from the top in figure 7 belongs to a spiral shaped wire, which was electrically driven by a constant electrical current, heating the whole wire up to 90°C. In this image, the spatial resolution was adjusted to be 64×64 pixels by controlling the pitch size of the XY scanner. The second image of figure 7 shows the thermal image of a human face, and the third image belongs to a cup full of 100°C water, both with a spatial resolution of 20×20 pixels. The right-hand images of figure 7 are the results of post processing of the left-hand images, performed in Python. For the post processing, a Gaussian filter was applied to the images followed by intensity adjustment and finally histogram equalization.

## 5. Conclusion

In this paper, thermal images belonging to the objects within the temperature range of 300 K to 400 K were captured by a fabricated single pixel superconducting TEB detector. The fabrication process of the associated superconducting thin film facilitates its application for large area substrates, suitable for arrayed structures with large number of integrated pixels. The detector was structurally designed to increase its sensitivity across the spectral range of interest. To further increase the sensitivity of the detector, the absorber layers were implemented on its backside. It has been shown that by employing a relatively low resolution imaging measurement, thermal images of acceptable quality could be reconstructed using modest post processing techniques.